\documentstyle{mn}
\def\degree{\nobreak\ifmmode{^\circ}\else{$^\circ$}\fi}
\def\arcmin{\nobreak\ifmmode{'}\else{$'$}\fi}
\def\arcsec{\nobreak\ifmmode{''}\else{$''$}\fi}
\def\hour{\nobreak\ifmmode{^{\rm h}}\else{$^{\rm h}$}\fi}
\def\minute{\nobreak\ifmmode{^{\rm m}}\else{$^{\rm m}$}\fi}
\def\second{\nobreak\ifmmode{^{\rm s}}\else{$^{\rm s}$}\fi}
\def\degreedot{\nobreak\ifmmode{^\circ\hskip-0.40em.\hskip0.08em}%
                         \else{$^\circ\hskip-0.40em.\hskip0.08em$}\fi}
\def\arcmindot{\nobreak\ifmmode{'\hskip-0.30em.\hskip0.02em}%
                         \else{$'\hskip-0.30em.\hskip0.02em$}\fi}
\def\arcsecdot{\nobreak\ifmmode{''\hskip-0.45em.\hskip0.08em}%
                         \else{$''\hskip-0.45em.\hskip0.08em$}\fi}
\def\seconddot{\nobreak\ifmmode{^{\rm s}\hskip-0.35em.\hskip0.05em}%
                         \else{$^{\rm s}\hskip-0.35em.\hskip0.05em$}\fi}
\def\spose#1{\hbox to 0pt{#1\hss}}
\def\simlt{\mathrel{\spose{\lower 3pt\hbox{$\mathchar"218$}}
     \raise 2.0pt\hbox{$\mathchar"13C$}}}
\def\simgt{\mathrel{\spose{\lower 3pt\hbox{$\mathchar"218$}}
     \raise 2.0pt\hbox{$\mathchar"13E$}}}
\def\OII{[O{\footnotesize~II}]}
\def\OIII{[O{\footnotesize~III}]}

\begin{document}
\title{A new look at the isotropy of narrow line emission in extragalactic
radio sources}
\author[C. Simpson]{Chris Simpson\\
Jet Propulsion Laboratory, California Institute of Technology, MS
169--327, 4800 Oak Grove Drive, Pasadena, CA 91109, USA}

\maketitle

\begin{abstract}
We undertake a quantitative investigation, using Monte Carlo simulations,
of the amount by which quasars are expected to exceed radio galaxies in
optical luminosity in the context of the `receding torus' model. We
compare these simulations with the known behaviour of the
\OIII~$\lambda$5007 and \OII~$\lambda$3727 emission lines and conclude
that \OIII\ is the better indicator of the strength of the underlying
non-stellar continuum.
\end{abstract}

\section{Introduction}

It is widely believed that radio-loud quasars and radio galaxies differ
from each other only in terms of the angle between their radio axes and
the line of sight (Scheuer 1987; Barthel 1989). Quasars are observed
fairly close to the line of sight ($\simlt 45\degree$) and therefore
frequently exhibit the effects of beaming, such as superluminal motion and
luminous flat-spectrum radio cores, whereas radio galaxies are observed
with their axes close to the plane of the sky, and do not show these
effects. However, quasars also possess a luminous non-stellar optical
continuum and broad emission lines, which are absent in radio galaxies. It
has therefore been proposed that there is material around the nucleus
which lies preferentially in the plane perpendicular to the radio axis and
obscures the central regions from view in radio galaxies. Because of the
geometry of this material, it is often referred to as the ``torus'', but
other geometries (e.g.\ a warped disk) are possible. Although the broad
lines are hidden from direct view by this material, the narrow line region
(NLR) is much larger in size and should be less strongly affected. It is
therefore possible that narrow line luminosity could be an isotropic
measure of the strength of the non-stellar continuum that is otherwise
invisible in radio galaxies.

Jackson \& Browne (1990) tested this idea by comparing the
\OIII~$\lambda$5007 luminosities of quasars and radio galaxies with
similar redshifts and extended radio luminosities. They reported that
quasars are 5--10 times more luminous in this line than radio galaxies,
and attributed the difference to higher extinction in radio galaxies. Hes,
Barthel \& Fosbury (1993) performed a similar test using the
\OII~$\lambda$3727 doublet and found that there was no measurable
difference between the line luminosities of the two classes. Since \OII\
is at a shorter wavelength than \OIII\ and would be more greatly affected
by a foreground screen of reddening, they suggested that most of the
\OIII\ emission is produced close to the nucleus and is still obscured by
the torus in radio galaxies, whereas \OII\ is produced farther out and is
unaffected.

Unfortunately, Jackson \& Browne's analysis was biased because, although
the radio galaxies were selected from the 3CR-based complete sample of
Laing, Riley \& Longair (1983; also called 3CRR), the quasars were drawn
from a number of incomplete surveys and were therefore subject to
uncertain selection effects, most notably the tendency to preferentially
include optically bright objects. Since line and continuum luminosities
are very well correlated in quasars, their result would be biased towards
finding systematically higher line luminosities in the quasars. In
addition, Jackson \& Browne's radio galaxy sample included some objects
with very weak emission lines (Class~B optical spectra; Hine \& Longair
1979) that it is now believed may not belong to the unified scheme. An
unbiased analysis significantly reduces the magnitude of Jackson \&
Browne's result, but the quasars are still about twice as luminous in
\OIII\ than the radio galaxies. This result is also seen in the higher
\OIII/\OII\ ratios in broad-line objects than in narrow-line objects
(e.g.\ Saunders et al.\ 1989; Tadhunter et al.\ 1993). At higher redshift,
however, the \OIII\ luminosities of the two classes of object are
comparable (Jackson \& Rawlings 1997).

Rawlings \& Saunders (1991) note that the tendency for quasars to have a
higher emission line luminosity than radio galaxies could be the result of
a classification bias. This could explain the discrepancy in \OIII\
luminosities between the two classes, but appears to run counter to the
similarity in the \OII\ line emission. In this {\it Letter\/}, we apply
the simple receding torus model (Lawrence 1991; Hill, Goodrich \& Depoy
1996) to provide a quantitative explanation of the difference in \OIII\
luminosities between quasars and radio galaxies (and its redshift
dependence), and explain why a similar effect is not seen in \OII. The
free parameters involved in our explanation are constrained by observed
quantities independent of the emission line luminosities. We conclude that
the luminosity of the \OIII~$\lambda$5007 line is an excellent indicator
of the strength of the underlying non-stellar continuum.

\section{Method}

\subsection{The receding torus model}

\begin{figure}
\vspace*{171pt}
\includegraphics{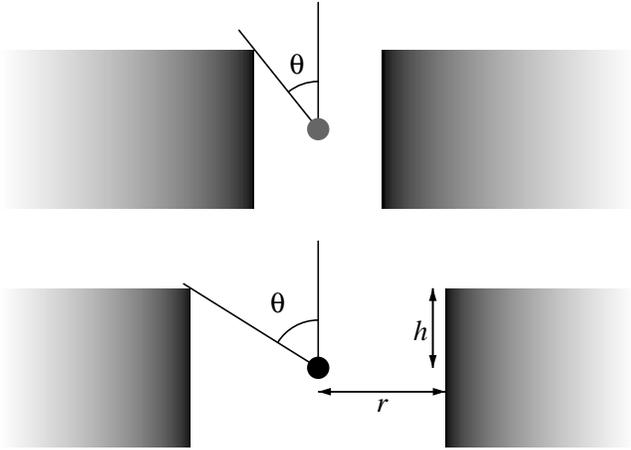}
\caption[]{Schematic representation of the receding torus model. The
opening angle, $\theta$, for the low luminosity object (top) is fairly
small. In the high luminosity object (bottom), the inner radius of the
torus, $r$, is larger, producing an increased opening angle provided the
half-height of the torus, $h$, remains constant.}
\label{fig:rtm}
\end{figure}

The receding torus model, as proposed by Lawrence (1991) and illustrated
in Figure~\ref{fig:rtm}, is fairly simple. Dust evaporates at a
temperature of about 1500\,K, and cannot survive closer to the nucleus
than the radius where the temperature of the nuclear radiation field is
hotter than this. All quasars display a near-infrared bump longward of
1\,$\mu$m that is believed to be caused by thermal emission from dust, and
the constancy of wavelength at which this bump appears indicates that dust
is always present at the hottest possible temperatures. The inner radius
of the torus is therefore determined by the radius at which dust
evaporates, which scales as $r \propto L^{0.5}$. If the half-height, $h$,
of the torus is independent of the source luminosity, then the half
opening angle $\theta = \tan^{-1} r/h$ will be larger in the more luminous
objects. Hill et al.\ (1996) find that this model fits the observed
nuclear extinctions in radio galaxies very well.

We are more likely to be viewing the luminous objects within their cones,
and are consequently more likely to classify them as quasars.
Qualitatively then, this simple model could explain the tendency for
quasars to have a higher \OIII\ luminosity than radio galaxies. We
therefore undertake a quantitative investigation using Monte Carlo
simulations.

\subsection{Simulations}

We consider an ensemble of objects with the same redshift and radio power,
which could reasonably be assumed to have similar torus properties (e.g.\
torus height). Their intrinsic optical--ultraviolet luminosities will have
a well-defined mean, $L_0$, determined by the strong correlation between
optical and radio luminosity, but will show scatter due to the intrinsic
dispersion in this relationship.  We assume this dispersion to be Gaussian
in $\log L$, with standard deviation $\sigma$. We define $\theta_0$ to be
the half opening angle for an object with the mean luminosity, $L_0$. An
object with luminosity $L$ with therefore have an half opening angle,
\[
\theta = \tan^{-1} [ (L/L_0)^{0.5} \tan \theta_0 ] ,
\]
and the probability that such an object will be observed as a radio galaxy
(i.e.\ our line of sight is outside the opening angle of the cone) is
\[
p_{\rm RG} = \cos \theta = (1 + \tan^2 \theta)^{-0.5} .
\]
Obviously, $p_{\rm QSO} \equiv 1 - p_{\rm RG}$.

\begin{figure}
\includegraphics{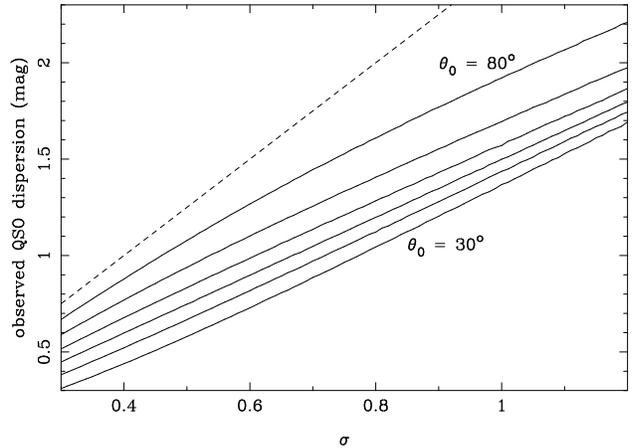}
\vspace*{60mm}
\caption[]{Observed dispersion in a radio-selected quasar sample as a
function of the dispersion in the parent AGN population. $\theta_0$ is the
half opening angle for the mean luminosity, $L_0$, and the solid lines are
for $\theta_0 = 30\degree$, 40\degree, 50\degree, 60\degree, 70\degree,
80\degree. The dashed line is the equality relationship.}
\label{fig:disp}
\end{figure}

For our ensemble of objects, we can therefore determine the ratio of the
mean optical--UV luminosities of quasars and radio galaxies (we evaluate
the mean in log space), for a given $\sigma$ and $\theta_0$.  However,
there exist observational data which allow us to fix these free
parameters. First, Serjeant et al.\ (1998) measure the scatter about the
mean optical--radio luminosity correlation for steep-spectrum quasars to
be 1.5\,magnitudes (0.6\,dex). This will be slightly smaller than
$\sigma$, since there should exist radio galaxies whose intrinsic optical
luminosities are on average fainter than those of the quasars. However,
Serjeant et al.'s sample was dominated by distant ($z>1$) quasars whose
high luminosities would imply a large opening angle for the torus. At
large opening angles, most objects are observed as quasars, and therefore
the dispersion in optical magnitude in a quasar sample is close to the
intrinsic dispersion of the parent AGN population (Figure~\ref{fig:disp}).
We thus adopt $\sigma = 0.6$. The second constraint is the observed quasar
(or radio galaxy) fraction; we must constrain $\theta_0$ such that the
fraction in our simulation matches that in a radio-selected sample.

\begin{figure}
\includegraphics{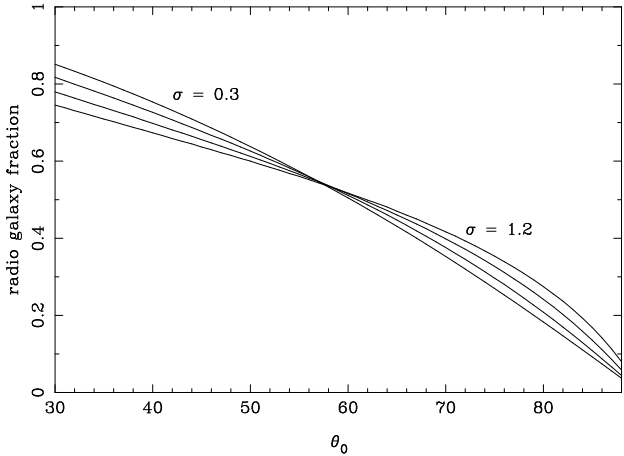}
\vspace*{60mm}
\caption[]{Observed radio galaxy fraction as a function of the torus
half opening angle, $\theta_0$, for an object with the mean luminosity,
$L_0$. The lines are for $\sigma = 0.3$, 0.6, 0.9, 1.2.}
\label{fig:frg}
\end{figure}

Following the reasoning of Laing et al.\ (1994) and others, when comparing
our results to observed quantities, we consider only those radio galaxies
with prominent emission lines. At $z < 0.43$ and $0\hour < \alpha <
13\hour$, where there exist high-quality optical spectra allowing
classification on the basis of broad H$\alpha$ (Laing et al.\ 1994), there
are 15 such radio galaxies and 9 quasars. The observed radio galaxy
fraction is therefore 0.625, corresponding to $\theta_0 = 50\degree$ for
$\sigma = 0.6$ (Figure~\ref{fig:frg}; the value is extremely insensitive
to the choice of $\sigma$).

\section{Results}

\subsection{Relative quasar/radio galaxy continuum and emission line
luminosities}

\begin{figure}
\includegraphics{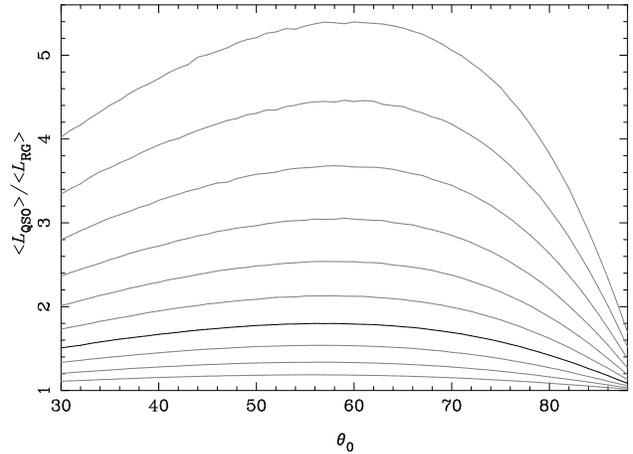}
\vspace*{60mm}
\caption[]{Ratio of mean quasar luminosity to mean radio galaxy luminosity
as a function of $\theta_0$ for various value of $\sigma$, the dispersion
in the optical--radio luminosity correlation of the parent population. The
lines run from $\sigma = 0.3$ (bottom) to $\sigma = 1.2$ (top) in steps of
0.1. The bold line is for $\sigma = 0.6$.}
\label{fig:rellum}
\end{figure}

Figure~\ref{fig:rellum} illustrates the amount by which quasars are
expected to be overluminous compared to radio galaxies as a function of
$\theta_0$ and $\sigma$, and for the values determined previously, we find
$\langle L_{\rm QSO} \rangle / \langle L_{\rm RG} \rangle = 1.8$. This is
effectively the same as the difference in \OIII\ luminosities for the two
groups of objects, which could therefore be explained if the \OIII\
emission line luminosity is directly proportional to the ionizing
luminosity. Its constancy of equivalent width over approximately three
orders of magnitude in luminosity (Miller et al.\ 1992) indicates that it
is.

\begin{figure}
\includegraphics{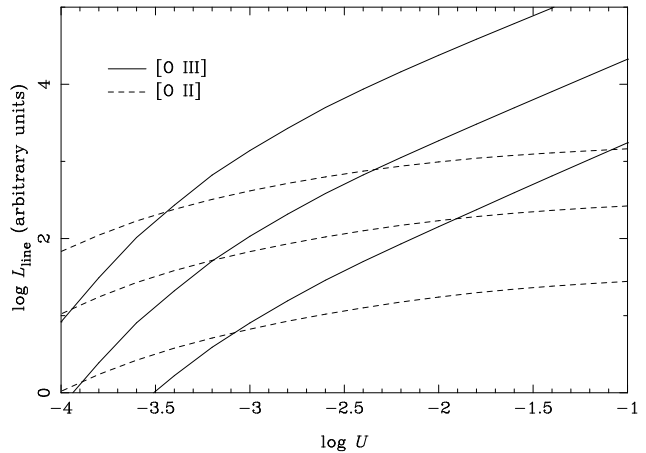}
\vspace*{60mm}
\caption[]{Luminosity of the \OIII\ (solid lines) and \OII\ (dashed lines)
lines as a function of ionization paramater, $U$ for an optically-thick
plane-parallel slab. Each set of curves is for values of the hydrogen
number density $n_{\rm H} = 10^2$, $10^3$, $10^4$\,cm$^{-3}$.}
\label{fig:linelum}
\end{figure}

Why then is a similar effect not seen in the \OII\ luminosity, since both
emission lines correlate well with optical luminosity over several orders
of magnitude in quasars? We believe that the answer is once again due to a
bias introduced by scatter. Although the effective ionization parameter of
the NLR does not correlate with luminosity (the range of quasar
luminosities is far larger than the observed range of $U$, $-3 \leq \log U
\leq -1$; e.g.\ Saunders et al.\ 1989), the most luminous objects at a
given redshift will also have the highest ionization parameters.
Figure~\ref{fig:linelum} shows the dependence of \OIII\ and \OII\
luminosity as a function of ionization parameter for an optically-thick
plane-parallel slab. This was calculated with the photoionization code
{\sc cloudy} Version 84.12a (Ferland 1993) using the AGN continuum
described in Mathews \& Ferland (1987), although effectively identical
results are obtained for power laws with a range of spectral index. Over
the range of effective ionization parameter typical of the NLR, the \OII\
luminosity is only very weakly dependent on the strength of the ionizing
continuum. Whereas a hundredfold increase in $U$ increases the luminosity
of \OIII\ by a factor of $\sim 200$, the \OII\ luminosity becomes only
$\sim 4$ times larger.  Although a single plane-parallel slab is obviously
an overly simplistic model for the NLR, there is very little density
dependence in the way the line luminosities vary with ionization
parameter.

\subsection{Redshift dependence}

The lessening of the effect at higher redshift is easy to understand,
since in the flux-limited 3CR sample this corresponds to higher luminosity
and therefore a larger mean opening angle. At such large angles,
variations in luminosity of an order of magnitude or less have little
effect on the angle, and hence the bias towards more luminous objects
being seen as quasars is smaller. This is apparent in
Figure~\ref{fig:rellum}. Although radio galaxies outnumber quasars 2:1 at
low $z$, at $z > 1$ the ratio is nearly reversed. In addition, there may
be many objects like 3C~22 (Economou et al.\ 1995; Rawlings et al.\ 1995)
or 3C~41 (Simpson, Rawlings \& Lacy 1998; Economou et al.\ 1998) that
possess broad H$\alpha$ which is difficult to observe from the ground at
$z \simgt 1.1$ due to it being redshifted into regions of poor atmospheric
transmission or the $H$ window, whose strong OH airglow lines hamper the
detection of broad features. The observed radio galaxy fraction of $\sim
40$\% at $z > 1$ should therefore be considered an upper limit, and it
provides a lower limit to $\theta_0 \simgt 70\degree$. This corresponds to
$\langle L_{\rm QSO} \rangle / \langle L_{\rm RG} \rangle < 1.7$, and the
ratio would drop below 1.5 if only one third of those objects currently
classified as radio galaxies are really quasars. It should also be noted
that measurement errors tend to be rather larger in near-infrared than in
optical spectroscopy, and so the study of Jackson \& Rawlings (1997) is
less sensitive to differences in the two classes than are low redshift
studies.

Jackson \& Rawlings (1997) do not offer any explanation as to why they
fail to observe a similar effect to that of Jackson \& Browne (1990), yet
it is naturally explained in our model as a luminosity, rather than a
redshift, dependence. Although our model might appear to predict that high
redshift samples whose luminosities are similar to those of low redshift
3CRR objects (e.g.\ a sample from the 7C catalogue) should also display a
separation in \OIII\ luminosity with broad/narrow line class, such a
prediction also depends on the assumption that the average torus height
does not vary with redshift since this also controls the mean opening
angle. We can predict that the difference between the \OIII\ luminosities
of radio galaxies and quasars should be a function of quasar fraction
(Figures~\ref{fig:frg} and \ref{fig:rellum}), and should effectively
vanish for samples with a large ($\simgt 80$\%) quasar fraction.

\subsection{Polarization properties}

Since our model requires that effectively all the \OIII\ emission is seen
directly, it appears to be in contradiction with the recent detection of
polarized \OIII\ in radio galaxies by di Serego Alighieri et al.\ (1997).
These authors calculate that 80--90\% of the total \OIII\ emission is
obscured in radio galaxies, but their calculations assume that the
observed featureless continuum is entirely scattered. It is known that
Seyfert galaxies possess an extended featureless continuum (FC2; Tran
1995) and it is reasonable to imagine that radio galaxies do also, thus
lowering the estimate of the obscured fraction. In addition, the \OIII\
emission may be polarized, even if none is obscured, if the scattering
region is distributed anisotropically. Although di Serego Alighieri et
al.\ disfavour this idea, it is not in conflict with observations, and
would be expected if the scattering material is associated with the
outflow. The fact that polarized \OIII\ emission is detected in 3C~227
despite our relatively clear view of the nucleus seems to indicate that
significant obscuration is not required for the line emission to be
polarized. We also note that the size of the \OIII-emitting region that
must be obscured is much larger than estimates for the height of the torus
derived from the dust evaporation radius and cone opening angle. While the
polarization data are intriguing, we therefore feel that our model is not
in conflict with them.

\section{Summary}

We have used a simple receding torus model, whose free parameters are
constrained by observation, to show that low redshift 3CRR quasars should
be, on average, about twice as luminous in their ionizing continua as
radio galaxies of the same radio luminosity. This difference should also
be seen in their \OIII, but not their \OII, emission line luminosities, in
agreement with observation. For samples with a higher quasar fraction,
such as the high redshift 3CRR objects, the difference in ionizing
luminosities between quasars and radio galaxies should be smaller, and
there should therefore be less of a difference in their \OIII\
luminosities, again in line with observation.

This model leads to the conclusion that the \OIII~$\lambda$5007 emission
line, and not the \OII~$\lambda$3727 doublet, is an unbiased indicator of
the intrinsic optical--ultraviolet luminosity of both quasars and radio
galaxies.

\section*{Acknowledgments}

The author would like to thank Peter Eisenhardt for a useful discussion
and Steve Rawlings for a critical reading of the manuscript. This work was
performed by the Jet Propulsion Laboratory, California Institute of
Technology, under a contract with NASA.

\end{document}